\documentclass[conference,a4paper]{IEEEtran}
\usepackage[cmex10]{amsmath}
\usepackage{amssymb,graphics,graphicx,color,float}
\usepackage{cite}

\def\oggi{\number\year-%
  \ifnum 10>\month 0\fi\number\month-%
  \ifnum 10>\day 0\fi\number\day}
	
\def\ie{{\it i.e.,\ }}

\newcommand{\E}{\mbox{${\mathbb{E}}$}}

\def\reals{ { {\rm  I \kern-0.15em R }  } }
\def\complex{\hbox{ \,{{\rm C} \kern-0.50em \raise0.20ex {  |}}\,}}

\def\Hbf{\mathbf H}
\def\Ibf{\mathbf I}

\def\Pbf{\mathbf P}

\def\Wbf{\mathbf W}

\def\abf{\mathbf a}

\def\ebf{\mathbf e}

\def\hbf{\mathbf h}

\def\wbf{\mathbf w}
\def\xbf{\mathbf x}
\def\ybf{\mathbf y}

\def\0bf{\mathbf 0}
\def\1bf{\mathbf 1}

\def\be{\vskip .3cm \begin{equation}}
\def\ee{\end{equation} \vskip .4cm \noindent}
\newcounter{assumpnr}
\renewcommand{\theassumpnr}{\arabic{assumpnr}}

\newcommand{\tr}[1]{\mbox{{$\mathrm{Tr}\left\{#1\right\}$}}}

\newcommand{\fro}[1]{\mbox{{$\left\|#1\right\|_{F}^2$}}}

\def\deg{$^\circ${ }}

\def\Pa0kp{\mbox{{$\Pbf^\perp_{\abf_{0k}}$}}}

\def\Pa0kp{\mbox{{$\Pbf^\perp_{\abf_{0k}}$}}}

\def\figpath{./}
\newcommand{\figl}{8cm}

\newcommand{\figs}{4.5cm}
\DeclareRobustCommand*{\IEEEauthorrefmark}[1]{\raisebox{0pt}[0pt][0pt]{\textsuperscript{\footnotesize #1}}}
\IEEEoverridecommandlockouts
\begin{document}
\title{Analysis of Massive MIMO With Hardware Impairments and Different Channel Models}

\vspace{-2cm}

% use a single column layout for the different affiliations.
% In case the affiliation is too long then use \\ to create multiple lines, as shown in affiliation 4
% Too long lines can result in margin upload warnings ....
%\author{\IEEEauthorblockN{
%Fredrik Athley\IEEEauthorrefmark{1}, Giuseppe Durisi\IEEEauthorrefmark{2}, Ulf Gustavsson\IEEEauthorrefmark{2} }}                                     % ...

%\IEEEauthorblockA{\IEEEauthorrefmark{1}% 1st affiliations
%Ericsson Research, Ericsson AB, Gothenburg, Sweden, E-mail: fredrik.athley@ericsson.com}

\author{\IEEEauthorblockN{
		Fredrik Athley\IEEEauthorrefmark{1},   % 1st author, 1st affiliations
		Giuseppe Durisi\IEEEauthorrefmark{2}\thanks{This work was partly supported by the Swedish Foundation for Strategic Research under grant SM13-0028.},   % 2nd author, 2nd affiliations
		Ulf Gustavsson\IEEEauthorrefmark{1}    % 3rd author, 3rd affiliations
	}                                     % ...
	%\\
	\IEEEauthorblockA{\IEEEauthorrefmark{1}% 1st affiliations
		Ericsson Research, Ericsson AB, Gothenburg, Sweden}
	\IEEEauthorblockA{\IEEEauthorrefmark{2}% 2nd affiliations
	Dept. of Signals and Systems, Chalmers University of Technology, Gothenburg, Sweden}
}

\maketitle

\begin{abstract}
Massive Multiple-Input Multiple-Output (MIMO) is foreseen to be one of the main technology components in next generation cellular communications (5G). In this paper, fundamental limits on the performance of downlink massive MIMO systems are investigated by means of simulations and analytical analysis. Signal-to-noise-and-interference ratio (SINR) and sum rate for a single-cell scenario multi-user MIMO are analyzed for different array sizes, channel models, and precoding schemes. The impact of hardware impairments on performance is also investigated. Simple approximations are derived that show explicitly how the number of antennas, number of served users, transmit power, and magnitude of hardware impairments affect performance.
\end{abstract}
%{\smallskip \keywords Massive MIMO, 5G, precoding, array antennas}

\IEEEpeerreviewmaketitle

%\nocite{*}

\vspace{7pt}
\section{Introduction}
Massive Multiple-Input Multiple-Output (MIMO) systems are foreseen to be one of the main technology components in next generation cellular communication systems (5G). The basic idea with massive MIMO is to use a large number of antenna elements at the base station (BS)---an order of magnitude larger than used in current systems and much larger than the number of concurrently served user equipment (UE)---in order to achieve high spatial resolution and array gain. The high spatial resolution enables high system capacity by spatial multiplexing of UEs. The array gain enables high energy efficiency. Furthermore, the averaging effects obtained by using a large number of antenna elements make the channel behave almost deterministically, which has the potential to simplify radio resource management. 

In order for massive MIMO to be economically feasible, the cost per antenna element and its associated radio and base band branches must be significantly less than in current systems. Therefore, the requirements put on such components must be less stringent than in current systems. Hence, massive MIMO can be seen as a paradigm shift from using a few expensive antennas to many cheap antennas~\cite{larsson14-02a}.

\paragraph*{Contributions} In this paper, a performance analysis of downlink massive MIMO is presented. The focus is on the impact of basic antenna and channel model parameters on system performance. Downlink signal-to-interference ratio (SINR) and sum rate for the matched filter\footnote{Sometimes also referred to as conjugate beamforming or maximum ratio transmission.} (MF) and zero forcing (ZF) precoders are computed for a single-cell scenario using line-of-sight (LoS), independent and identically distributed (IID) Rayleigh, and statistical ray-based channel models. Simulation results are presented together with simple analytical expressions where the throughput dependence on the system parameters is transparent. Furthermore, the impact of phase and amplitude errors in the precoding weights is analyzed.

\paragraph*{Previous works} A lower bound on the massive MIMO downlink performance has been derived in~\cite{yang13-02b}, under the assumption that the propagation channel is learnt through the transmission of orthogonal pilot sequences in the uplink.
Hardware impairments in massive MIMO have been the subject of several recent theoretical investigations.
Most of these works model the hardware impairments as power-dependent Gaussian additive noise, which simplifies the throughput analyses~\cite{bjornson14-01a}.
More realistic multiplicative hardware models have been adopted in \emph{e.g.},~\cite{Gustavsson2014}.
Comparisons between the downlink throughput corresponding to measured propagation channels and ideal channel models have been reported in \emph{e.g.},~\cite{gao12-11a}.

\vspace{7pt}
\section{System Model}
We consider a single cell with no external interference. The downlink signals received by $K$ co-scheduled UEs, each with a single antenna, served by a BS  with $M$ antenna elements are modeled by the following equation
\begin{eqnarray}
\label{eq:signalmodel}
\ybf=\sqrt{P}\Hbf\Wbf\xbf + \ebf.
\end{eqnarray}
Here, $\ybf=[\begin{array}{ccc}
y_{1} & \cdots & y_{K} 
\end{array}]^T$ is a $K\times 1$ vector containing the signals received by the UEs, $P$ is the BS  transmit (Tx) power and
$\Hbf=[\begin{array}{ccc}
\hbf_{1}^T & \cdots & \hbf_{K}^T 
\end{array}]^T$ 
is the $K\times M$ channel matrix, with $\hbf_{k}$ being the $1\times M$ channel vector from the $M$ BS antenna elements  to UE $k$. Furthermore, $\Wbf=[\wbf_1 \cdots \wbf_K]$ denotes the $M\times K$ precoding matrix, $\xbf$ the $K\times 1$ vector of transmitted symbols, and $\ebf$ a $K\times 1$ zero-mean complex Gaussian vector with covariance matrix $N_0\Ibf$ modeling the receiver noise in the UEs. 

When increasing the number of antenna elements in the analysis, the Tx power will be scaled as $1/M$ in order to compensate for the increased array gain. As a result, a constant signal-to-noise ratio (SNR) operating point is maintained. More precisely, the Tx power for a given number of antenna elements is set so that a certain target SNR, $\mathrm{SNR}_t$, is obtained in the interference-free (IF) case, \ie
\begin{eqnarray}
\label{eq:Pnorm}
P=N_0\mathrm{SNR}_t/M.
\end{eqnarray}

The BS antenna array is assumed to be a uniform linear array (ULA) in the horizontal plane with 0.6$\lambda$ element separation, where $\lambda$ is the wavelength. The radiation pattern of a single element is modeled according to \cite{3GPP:25996} with 90\deg azimuth half-power beam width. Mutual coupling between array elements is ignored. The UE is assumed to have a single isotropic antenna. Polarization is not modeled.

\vspace{7pt}
\section{Perfect Channel State Information}
In this section, the system throughput is analyzed assuming that the BS has perfect channel state information (CSI). The SINR and sum rate are calculated for MF and ZF precoding schemes. Some analytical results are first derived for an IID Rayleigh channel model and then compared with other channel models by means of simulations.

The SINR for UE $k$ for a given channel realization $\Hbf$ is given by
\begin{eqnarray}
\gamma_k=\frac{P|\hbf_k\wbf_k|^2}{P\sum_{j=1 \atop j\neq k}^K|\hbf_k\wbf_j|^2 + N_0 }.
\end{eqnarray}
The maximum achievable rate for UE $k$ is then given by
\begin{eqnarray}
\label{eq:exact_rate}
R_k=\E\left[\log_2\left(1+\gamma_k\right)\right]
\end{eqnarray}
where the expectation is taken over the channel realizations. The sum rate is obtained by summing the rates of all concurrently served UEs according to 
\begin{eqnarray}
\label{eq:sum_rate}
R = \sum_k R_k.
\end{eqnarray}

\subsection{IID Rayleigh}
In this section, simple approximations of the rate in \eqref{eq:exact_rate} are derived under the assumption of an IID Rayleigh channel, \ie a channel for which $\Hbf$ has IID zero-mean complex Gaussian elements. The expectation in \eqref{eq:exact_rate} is difficult to compute analytically. However, as will be shown by simulations later in this paper, the following approximation is accurate
\begin{eqnarray}
\label{eq:rate}
R_k\approx\log_2(1+\mathrm{SINR}_k),
\end{eqnarray}
where
\begin{eqnarray}
\label{eq:SINR}
\mathrm{SINR}_k=\frac{P\E[|\hbf_k\wbf_k|^2]}{P\sum_{j=1 \atop j\neq k}^K\E[|\hbf_k\wbf_j|^2] + N_0 }.
\end{eqnarray}
Analytical approximations of the SINR for MF and ZF precoders are derived below. These can be used together with \eqref{eq:sum_rate}--\eqref{eq:rate} to estimate the sum rate in the cell.

\subsubsection{MF}
The matched filter precoding vector for UE $k$ is
\begin{eqnarray}
\wbf_k=\hbf_k^H/\lVert\hbf_k\rVert,
\end{eqnarray}
where the normalization by $||\hbf_k||$ assures the same transmitted power in all channel realizations and equal power allocation to all UEs.
The expectations in \eqref{eq:SINR} are then given by 
\begin{eqnarray}
\E[|\hbf_k\wbf_k|^2]=\E\left[\left|\hbf_k\hbf_k^H/\lVert\hbf_k\rVert\right|^2\right]=\E[\lVert\hbf_k\rVert^2]=M,
\end{eqnarray}
and
\begin{eqnarray}
\E[|\hbf_k\wbf_j|^2]=\E\left[\left|\hbf_k\hbf_j^H/\lVert\hbf_j\rVert\right|^2\right]=1.
\end{eqnarray}
Hence,
\begin{eqnarray}
\label{eq:SINRMF}
\mathrm{SINR}_k=\frac{PM}{P(K-1)+ N_0 }.
\end{eqnarray}
Using the power normalization in \eqref{eq:Pnorm} we obtain
\begin{equation}
\mathrm{SINR}_k=\frac{\mathrm{SNR}_t}{1+\mathrm{SNR}_t(K-1)/M}.
\end{equation}

\subsubsection{ZF}
With equal power allocation to all UEs, the ZF precoding vector is 
\begin{eqnarray}
\label{eq:WZF}
\wbf_{k}=(\Hbf^\dag)_k/\sqrt{\fro{(\Hbf^\dag)_k}},
\end{eqnarray}
where $(\Hbf^\dag)_k$ denotes the $k$-th column of $\Hbf^\dag$, $\lVert\cdot\rVert_F$ the Frobenius norm, and $\Hbf^\dag = \Hbf^H(\Hbf\Hbf^H)^{-1}$. In order to simplify the analytical calculations, the ZF precoding matrix is approximated by 
\begin{equation}
\label{eq:WcH}
\Wbf = c\Hbf^\dag,
\end{equation}
where $c$ is a normalization constant obtained by solving
\begin{equation}
\label{eq:ZFconstraint}
\E[\fro{\Wbf}]=K.
\end{equation}
This normalization makes the average transmitted total power to all $K$ UEs equal to $K$, but does not guarantee same power in all channel realizations and equal power allocation to all UEs. However, for an IID Rayleigh channel the difference turns out to be small when the number of antenna elements is large.

Under the assumption of an IID Rayleigh channel, $\Hbf\Hbf^H$ is a $K\times K$ central complex Wishart matrix with $M$ degrees of freedom and covariance matrix equal to the identity matrix.
It follows then that \cite[p.~26]{Tulino2004}
\begin{equation}
\E\left[\tr{(\Hbf\Hbf^H)^{-1}}\right]=K/(M-K)
\end{equation}
and the normalization constant is obtained according to
\begin{equation}
c^2\E\left[\fro{\Hbf^\dag}\right]=K \; \Rightarrow \; c=\sqrt{M-K}.
\end{equation}
The received signal according to \eqref{eq:signalmodel} is then given by
\begin{equation}
\ybf=\sqrt{P(M-K)}\Hbf\Hbf^\dag\xbf+\ebf=\sqrt{P(M-K)}\xbf+\ebf.
\end{equation}
Hence, 
\begin{equation}
\label{eq:SINRZF}
\mathrm{SINR}_k=P(M-K)/N_0=\mathrm{SNR}_t(1-K/M).
\end{equation}

\subsubsection{Simple rule of thumb}
The SINR approximations in \eqref{eq:SINRMF} and \eqref{eq:SINRZF} can be used to derive a simple rule of thumb on how many antennas are needed to reach a certain performance for a given number of co-scheduled UEs. For example, the number of antennas needed to reach an SINR that is 3 dB away from the IF SINR, $\mathrm{SNR}_t$, can be obtained by setting $\mathrm{SINR}=\mathrm{SNR}_t/2$. This leads to the following number of antennas for MF
\begin{equation}
\label{eq:RoTMF}
M=(K-1)\mathrm{SNR}_t,
\end{equation}
and for ZF
\begin{equation}
\label{eq:RoTZF}
M=2K.
\end{equation}

\subsubsection{Comparison between analytical and simulation results}
\begin{figure}[h]
	\centerline{\includegraphics[width=\figl]{\figpath/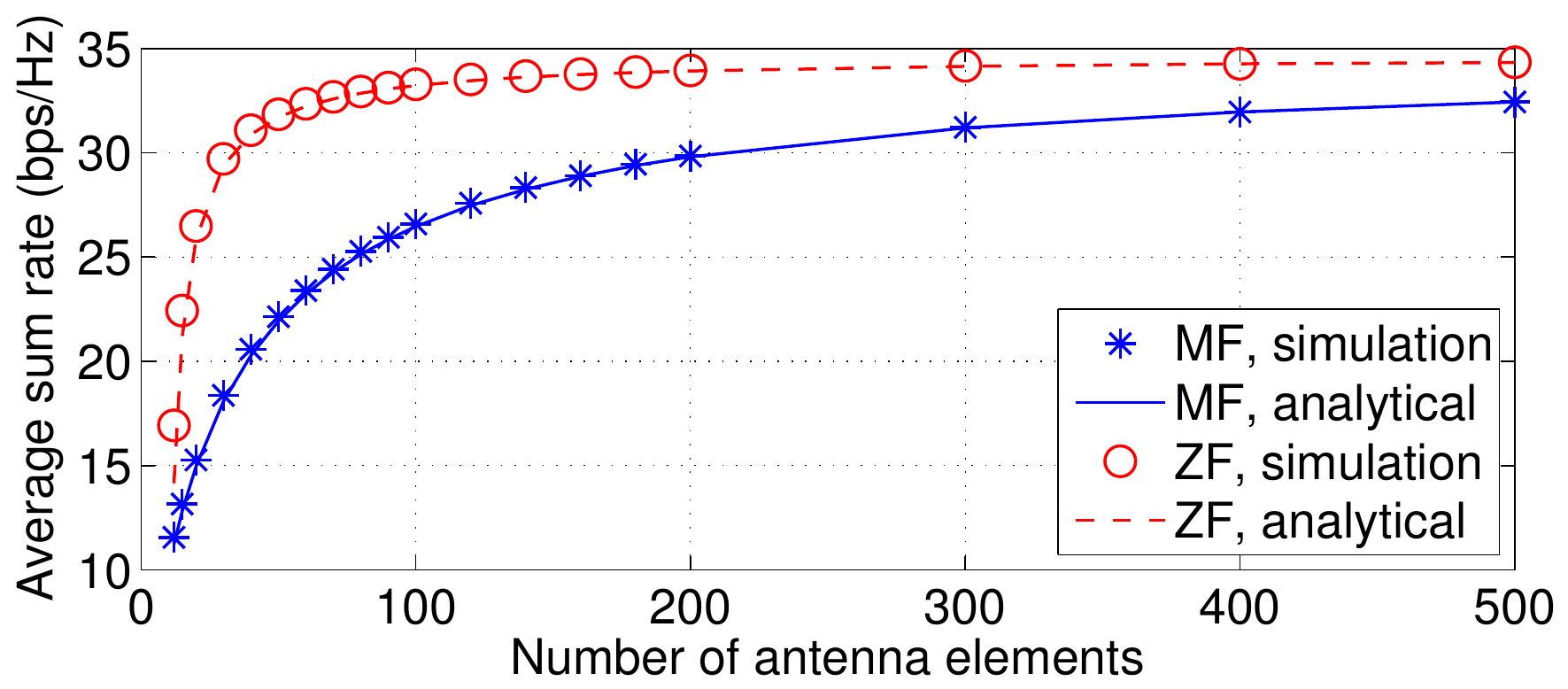}}
	\caption{Average sum rate vs. number of antennas for an IID Rayleigh channel using MF and ZF precoders. Comparison between analytical and simulation results with $K=10$ and $\mathrm{SNR}_t=10$ dB.}
	\label{fig:analytical_vs_sim_no_errors}
\end{figure}

Figure~\ref{fig:analytical_vs_sim_no_errors} shows a comparison of the analytical approximations in \eqref{eq:rate}--\eqref{eq:SINR}, \eqref{eq:SINRMF}, and \eqref{eq:SINRZF} with the the exact rate expression in \eqref{eq:exact_rate} obtained by Monte Carlo simulations.\footnote{In the Monte Carlo simulations of ZF performance, the precoding vector in \eqref{eq:WZF} has been used.} The plot shows the average sum rate vs. number of antenna elements for $K=10$ and $\mathrm{SNR}_t=10$ dB. Clearly, there is excellent agreement between the analytical and simulation results, also for moderately sized arrays. These results provide empirical support to the approximations made in deriving the analytical expressions. 

\subsection{Comparison of Channel Models}
The IID Rayleigh model is reasonable  when there is rich scattering around the BS and the UEs. However, in many cases the environment around the BS is such that there is spatial correlation among the BS antenna elements. We investigate the impact of spatial correlation on massive MIMO by using two types of correlated channel models:
\subsubsection{LoS channel model} We assume that there is a single, free-space planar wavefront from the BS to each UE. For a given azimuth angle of departure (AoD), the model is purely deterministic. However, the AoDs to different UEs are drawn from a uniform distribution over the interval [-60$^\circ$, 60$^\circ$]. 
\subsubsection{Statistical ray-based channel models} Two different ray-based models are investigated: the 3GPP spatial channel model (SCM) \cite{3GPP:25996} and the ITU urban macro (UMa) model \cite{ITU:2135} with indoor UEs added as described in \cite{3GPP:36819}.

%\begin{itemize}
%	\item Line-of-sight (LoS) channel model. In this model we assume that there is a single, free-space planar wavefront from the BS to each UE. For a given azimuth angle of departure (AoD), the model is purely deterministic. However, the AoDs to different UEs are drawn from a uniform distribution over the interval [-60$^\circ$, 60$^\circ$]. 
%\item Statistical ray-based channel models. Two different ray-based models are investigated: the 3GPP spatial channel model (SCM) \cite{3GPP:25996} and the ITU urban macro (UMa) model \cite{ITU:2135} with indoor UEs added as described in \cite{3GPP:36819}.
%\end{itemize} 
%

The IID Rayleigh and LoS channel models represent corner cases in terms of angular spread, with the IID model being spatially white and the LoS model having zero angular spread. The ray-based models lie in between with a mean angular spread of 15\deg in the SCM model and 14\deg and 26\deg mean angular spread for LoS and non-LoS UEs, respectively, in the ITU UMa model.

The SCM and ITU channel models also include models for path loss. In order to isolate the impact of path loss and spatial correlation on massive MIMO performance and to be able to compare with the unit gain IID Rayleigh channel, the path loss is first removed from channel matrices generated by the SCM and ITU models.

\begin{figure}[h]
	\centerline{\includegraphics[width=\figl]{\figpath/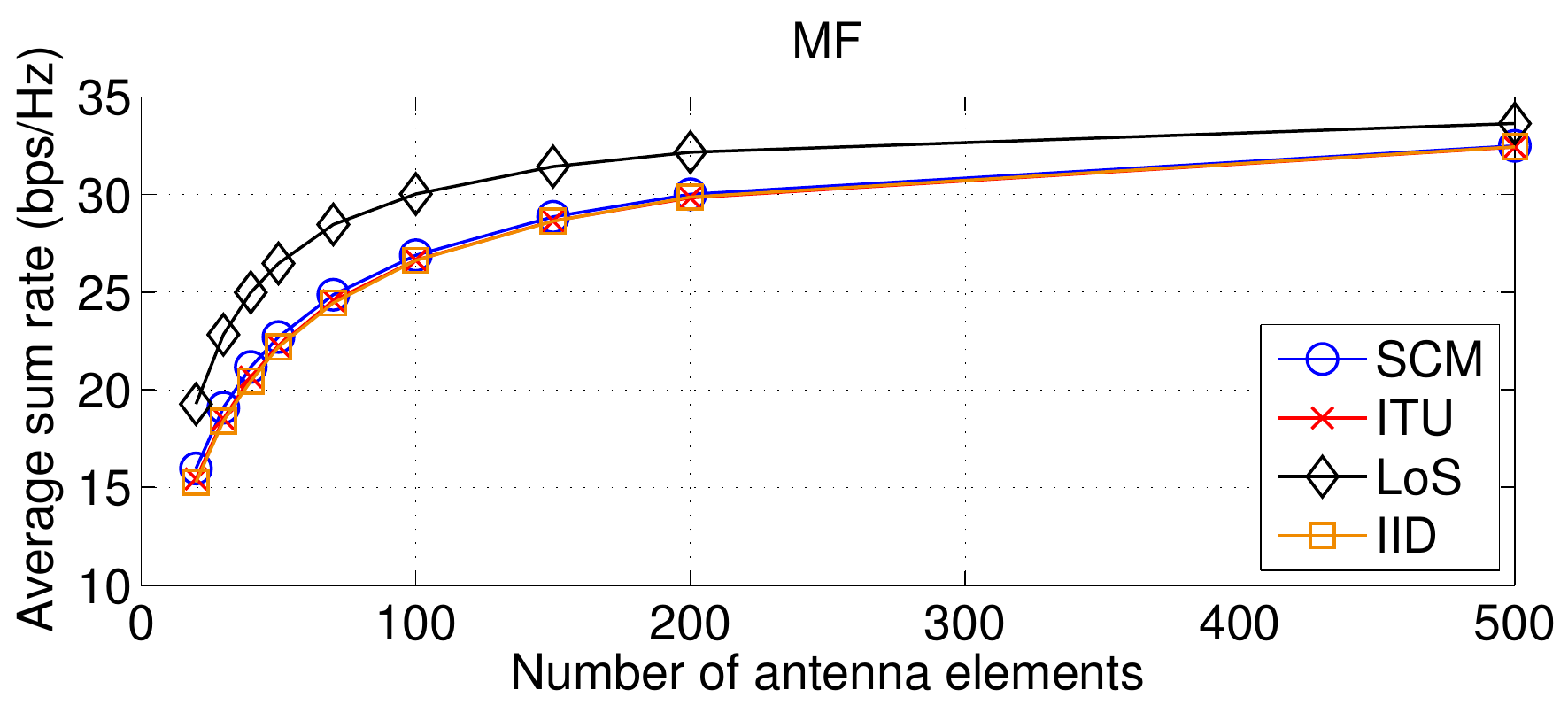}}	\centerline{\includegraphics[width=\figl]{\figpath/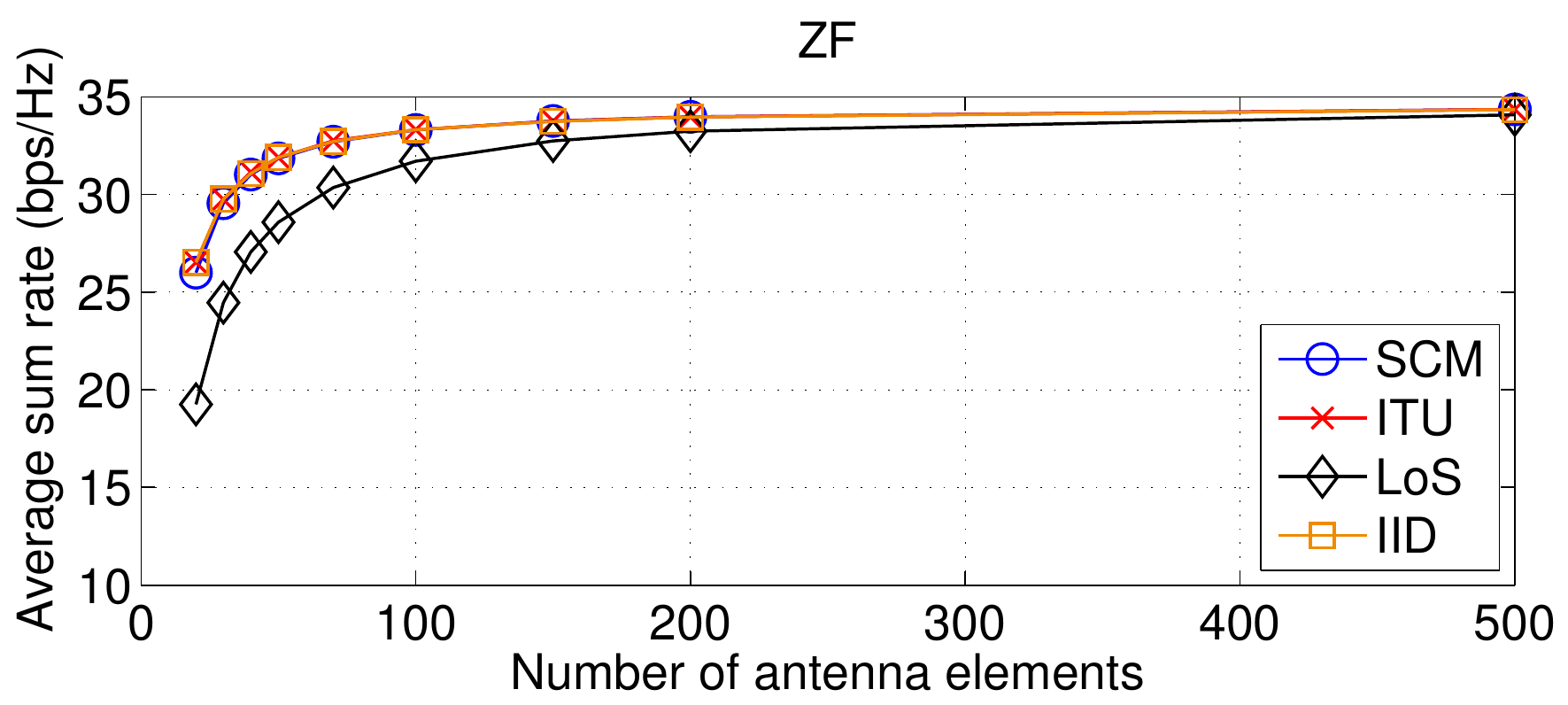}}
	\caption{Average sum rate vs. number of antenna elements for different channel models, top: MF, bottom: ZF.}
	\label{fig:channel_models}
\end{figure}

Figure~\ref{fig:channel_models} shows a comparison of performance with the different channel models when the channel gain has been normalized to one. With MF, the best performance is achieved with a LoS channel, while the LoS channel gives the worst performance when using a ZF precoder. The MF behavior can be explained by the low sidelobes attainable by a large antenna array, which result in good interference suppression in a LoS channel. The cause of the ZF behavior is that there is a gain penalty when two UEs are separated by less than a beam width in a LoS channel. Placing a null in the direction of one UE will, in this case, give a large gain drop to the other UE. In an uncorrelated channel, however, it is unlikely that two channel vectors are almost parallel. 

The other channel models give very similar performance for both the MF and ZF precoders. It is interesting to note that the SCM and ITU channel models give almost identical performance as an IID Rayleigh channel. A possible explanation to this is that different UEs will obtain different realizations of the channel rays. The only correlation between UEs is via the correlation between the large-scale parameters, such as angular spread, delay spread, and shadow fading. User-common clusters are not captured by these models. This is something that may be important when evaluating multiuser MIMO performance. 

\begin{figure}[h]
	\centerline{\includegraphics[width=\figl]{\figpath/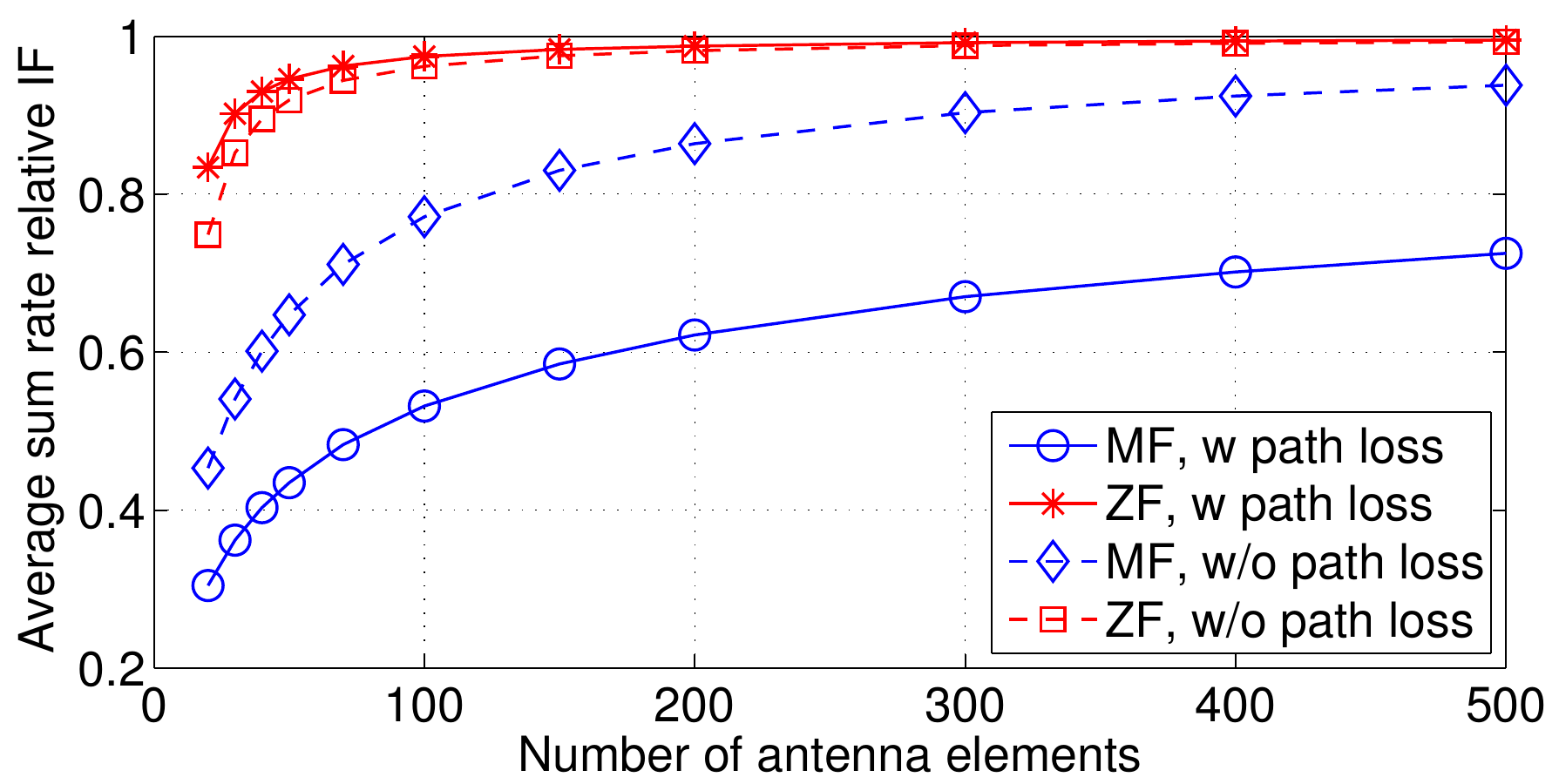}}	
	\caption{Average sum rate relative IF vs. number of antennas for different precoders with the ITU UMa channel with and without path loss}
	\label{fig:pathgainimpact}
\end{figure}

In order to investigate the impact of path loss difference between  UEs on performance, Figure~\ref{fig:pathgainimpact} shows the average sum rate relative the corresponding IF case as a function of the number of antenna elements using the ITU UMa model with and without path loss. The results show that MF relative performance is reduced significantly when using path loss in the channel model, while the impact on ZF relative performance is weak.

\section{Imperfect Channel State Information}
Perfect CSI at the BS cannot be achieved in reality so the results in the previous section should be interpreted as upper bounds on performance. In this section, we analyze the impact of imperfect CSI  using a simple model of hardware impairments. We model imperfect CSI  as a phase and amplitude error applied to the true channel. More specifically, the channel used in the downlink transmission is modeled as
\begin{equation}
\label{eq:errormodel}
\tilde{h}_{km}=(1+a_{m})e^{j\phi_m}h_{km}=\epsilon_mh_{km}
\end{equation}
where $\tilde{h}_{km}$ is the perturbed downlink channel coefficient between user $k$ and antenna $m$, $h_{km}$ is the 
corresponding true channel coefficient, and  $\epsilon_m\triangleq(1+a_{m})e^{j\phi_m}$. The amplitude error, $a_{m}$, and phase error, $\phi_m$, are assumed to be 
independent, zero-mean Gaussian random variables with variances $\sigma_a^2$ and $\sigma_\phi^2$, respectively. In the remainder of the paper, we shall refer to $\sigma_a$ as amplitude error, and measure it in dB, and to $\sigma_\phi$ as phase error, and measure it in degrees.

The adopted hardware impairment model is not a model of a particular component. Rather, it captures the aggregated effect of all errors in the system.  The polar form of this aggregated impairment model is supported by the fact that two of the largest impairments, \ie the power amplifier distorsion \cite{Pedro_Maas05} and the oscillator phase noise \cite{HajimiriPN-98} are multiplicative. It has been shown in \cite{Gustavsson2014} that system performance predictions based on this model are in good agreement with the ones based on more refined hardware models.

\setcounter{subsubsection}{0}
\subsubsection{MF}
In the case of errors, the MF precoding vector is given by the perturbed channel vector
\begin{eqnarray}
\wbf_k=\tilde{\hbf}_k^H/\lVert\tilde{\hbf}_k\rVert.
\end{eqnarray}
To simplify calculations, we approximate the norm by its expected value and instead use the following precoding vector
\begin{eqnarray}
\label{eq:wi}
\wbf_k=\tilde{\hbf}_k^H/\sqrt{\E[\lVert\tilde{\hbf}_k\rVert^2]}.
\end{eqnarray}
As we shall later show, this is an accurate approximation for large $M$. Using \eqref{eq:errormodel} we obtain
\begin{equation}
\E[||\tilde{\hbf}_k||^2]=\E[\sum_{m=1}^M|(1+a_{m})e^{j\phi_m}h_{km}|^2]=M(1+\sigma_a^2),
\end{equation}
and
\vspace{-5pt}
\begin{eqnarray}
\label{eq:Ehhtilde}   \lefteqn{\E\left[|\hbf_k\tilde{\hbf}_k^H|^2\right]=\E\left[|\sum_{m=1}^Mh_{km}h_{km}^*\epsilon_m|^2\right]=} \nonumber\\
&=& \E\Bigg[\sum_{m=1}^M|h_{km}|^4|\epsilon_m|^2 +\sum_{m=1}^M\sum_{n=1 \atop n\neq m}^M |h_{km}|^2|h_{kn}|^2\epsilon_m\epsilon_n^*\Bigg] \nonumber.
\end{eqnarray}
Since  $\E[|z|^4]=2\sigma^4$  for a complex Gaussian random variable $z$ with variance $\sigma^2$, we have that   $\E[|h_{im}|^4]=2$. To compute $\E[\epsilon_m]$, we use that the characteristic function, defined as $\psi(s)=\E[\exp(jsX)]$, of a Gaussian random variable $X$ with mean $\mu$ and variance $\sigma^2$ is given by $\psi(s)=e^{j\mu s-\sigma^2s^2/2}$. By setting $s=1$, we get
\begin{equation}
\E[e^{jX}]=e^{j\mu-\sigma^2/2}.
\end{equation}
Since, by assumption, $\phi$ is a zero-mean Gaussian random variable with variance $\sigma_\phi^2$, we obtain
\begin{equation}
\E[\epsilon_m]=\E[1+a_m]\, \E[e^{j\phi_m}]=e^{-\sigma_\phi^2/2}.
\end{equation}
The expectation in \eqref{eq:Ehhtilde} then simplifies to
\begin{equation}
\E[|\hbf_k\tilde{\hbf}_k^H|^2]=2M(1+\sigma_a^2)+M(M-1)e^{-\sigma_\phi^2}
\end{equation}
which, for large $M$, can be approximated by
\begin{equation}
\E[|\hbf_k\tilde{\hbf}_k^H|^2]\approx M^2e^{-\sigma_\phi^2}.
\end{equation}
Finally,
\begin{eqnarray}
\E[|\hbf_k\tilde{\hbf}_j^H|^2]=\E\left[\left\lvert\sum_{m=1}^Mh_{im}h_{jm}^*\epsilon_m\right\rvert^2\right]=M(1+\sigma_a^2).
\end{eqnarray}
The SINR approximation in \eqref{eq:SINR} is then given by
\begin{equation}
\label{eq:SINRerror}
\mathrm{SINR}_k=\frac{e^{-\sigma_\phi^2}}{1+\sigma_a^2}\frac{\mathrm{SNR}_t}{1+\mathrm{SNR}_t(K-1)/M}.
\end{equation}
Hence, the error-free SINR in \eqref{eq:SINRMF} is reduced by a factor $\exp(-\sigma_\phi^2)/(1+\sigma_a^2)$ in the presence of phase and amplitude errors. The factor $\exp(-\sigma_\phi^2)$ reflects that a phase error causes the ideal and perturbed channel vectors not to be parallel. An amplitude error does not destroy the alignment, but yields a gain reduction by a factor $1/(1+\sigma_a^2)$ due to the weight normalization in  \eqref{eq:wi} which is needed to assure conservation of energy.
If the errors are small, the error factor in \eqref{eq:SINRerror} can be approximated by Taylor expansions according to
\begin{equation}
\label{eq:error_factor}
\frac{e^{-\sigma_\phi^2}}{1+\sigma_a^2}\approx\frac{1-\sigma_\phi^2}{1+\sigma_a^2}\approx\frac{1}{1+\sigma_a^2+\sigma_\phi^2}=\frac{1}{1+\sigma^2}
\end{equation}
where $\sigma ^2=\sigma_a^2+\sigma_\phi^2$ is the total error variance. Hence, for small errors, the SINR degradation depends only on the sum of the variances of the phase and amplitude errors. The factor in \eqref{eq:error_factor} is the same as the gain reduction caused by phase and amplitude errors in phased arrays \cite{Mailloux2005}. Note that this SINR reduction does not depend on the number of antenna elements. Hence, it remains even in the limit $M\rightarrow\infty$. However, with the Tx power normalization used in this paper, the system will asymptotically be noise limited; so the asymptotic SINR loss can be compensated for by an increase in Tx power.

\subsubsection{Simple rule of thumb}
A rule of thumb for the required number of antennas to be 3 dB from the IF SINR is easily derived also in the case of phase and amplitude errors. Using the approximations in \eqref{eq:SINRerror} and \eqref{eq:error_factor}, the following expression is obtained
\begin{equation}
\label{eq:RoTMF_errors}
M=\frac{1+\sigma^2}{1-\sigma^2}(K-1)\mathrm{SNR}_t.
\end{equation}
Hence, the required number of antennas is increased by a factor $(1+\sigma^2)/(1-\sigma^2)$ compared to the error-free case in~\eqref{eq:RoTMF}. 

\subsubsection{Simulation results}
\begin{figure}[h]
	\centerline{\includegraphics[width=\figl]{\figpath/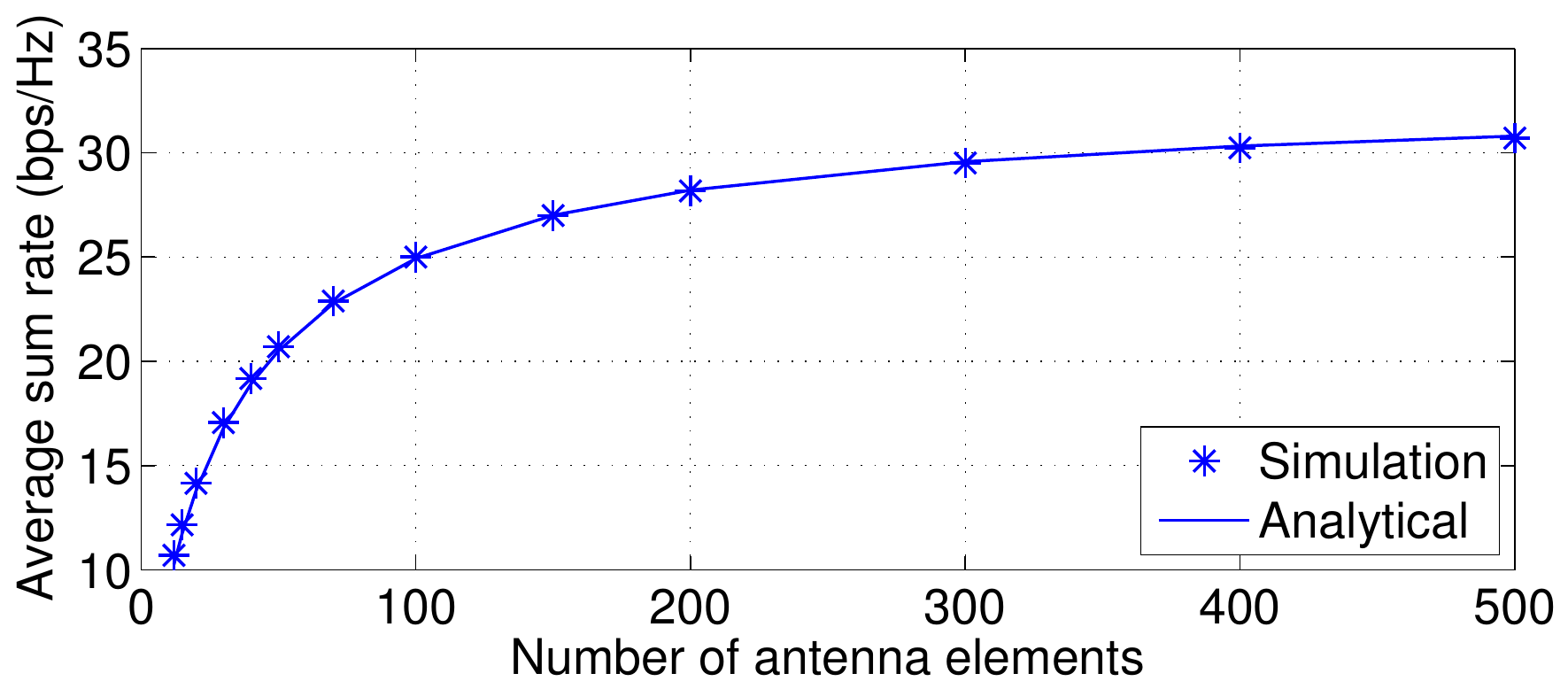}}
	\caption{Average sum rate vs. number of antennas for an IID Rayleigh channel using an MF precoder with 1 dB amplitude and 20$^\circ$ phase errors . Comparison between analytical and simulation results. $K=10$ and $\mathrm{SNR}_t=10$ dB.}
	\label{fig:analytical_vs_sim_errors}
\end{figure}

Figure~\ref{fig:analytical_vs_sim_errors}  shows a comparison between simulation results using an IID Rayleigh channel model and the approximations \eqref{eq:SINRerror}--\eqref{eq:error_factor} when the standard deviation of the phase and amplitude errors is 20$^\circ$ and 1 dB, respectively. The agreement between simulation and analytical results is excellent.
\vspace{-7pt}
Since analytical expressions for imperfect CSI have been derived only for the MF precoder some simulation results including also the ZF precoder are now presented. Since the IID, SCM, and ITU UMa channel models give very similar results only results for the IID model are given here.

\begin{figure}[h]
	\centerline{\includegraphics[width=\figs]{\figpath/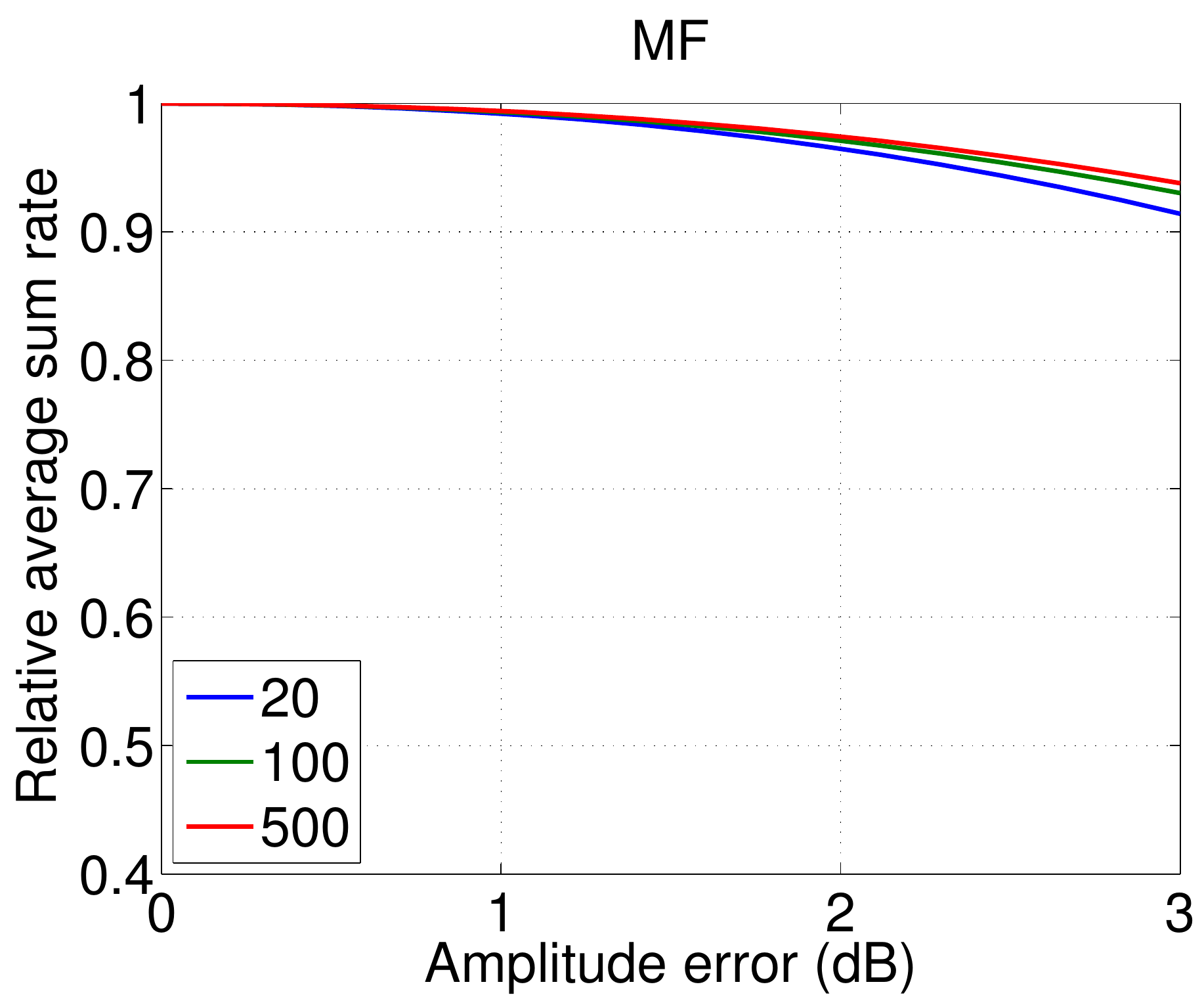}\includegraphics[width=\figs]{\figpath/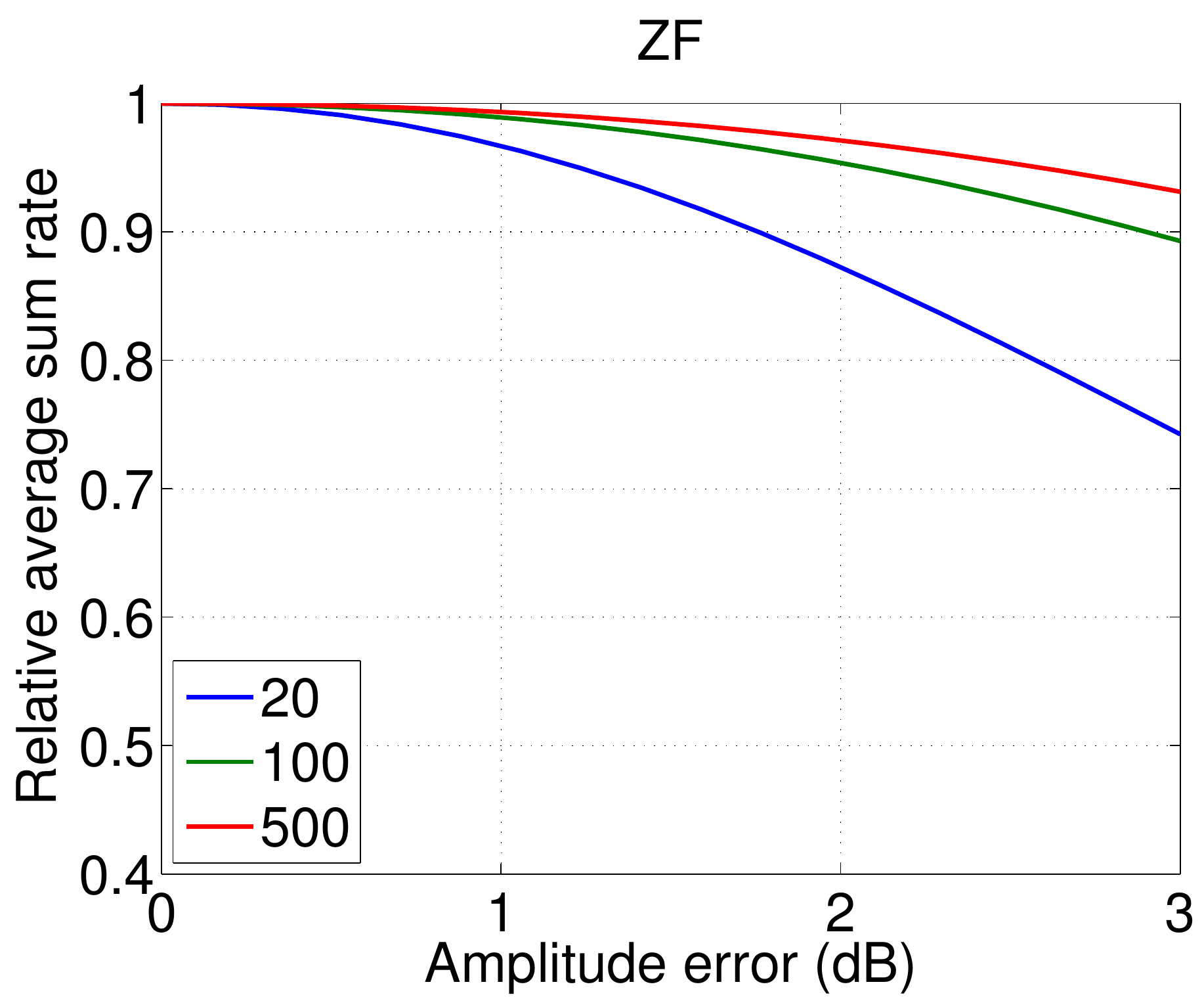}}
	\centerline{\includegraphics[width=\figs]{\figpath/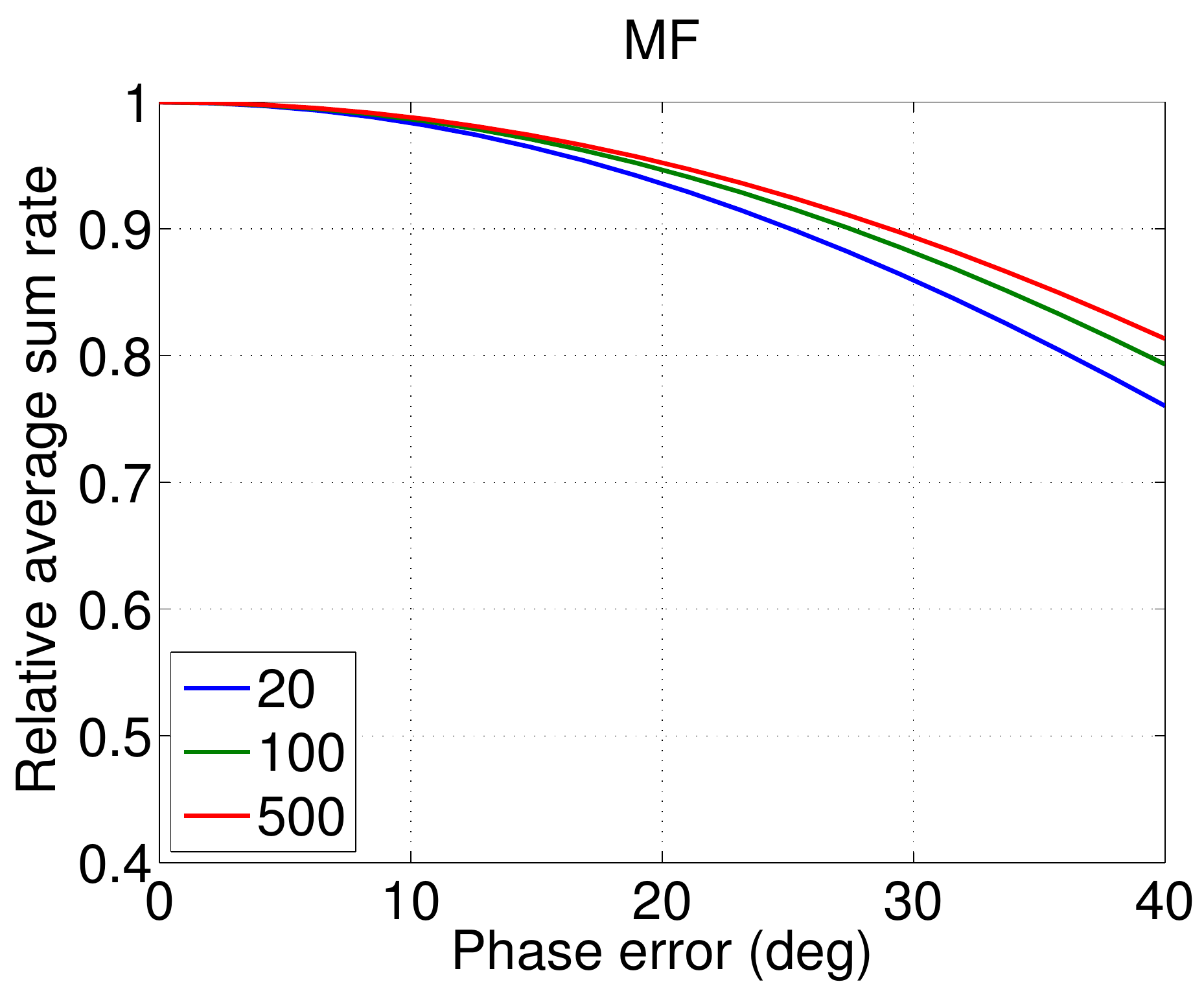}\includegraphics[width=\figs]{\figpath/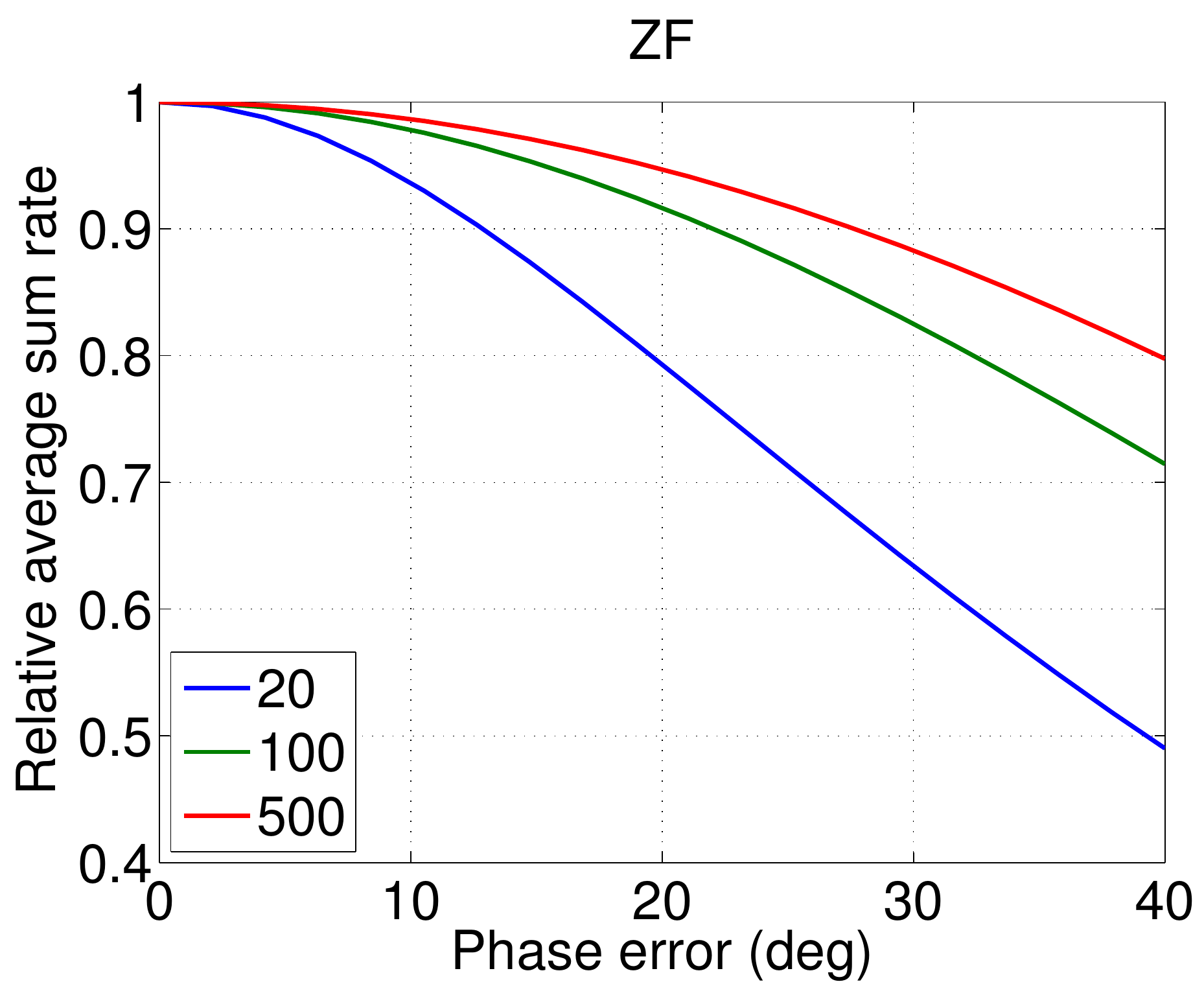}}
	\caption{Average sum rate relative the error-free case vs. amplitude (top) and phase (bottom) error for MF (left) and ZF (right) precoders. $K=10$ and $\mathrm{SNR}_t=10$ dB\vspace{-14pt}.}
	\label{fig:sim_ampl_phase_error}
\end{figure}

Figure~\ref{fig:sim_ampl_phase_error} shows average sum rate relative the error-free case vs. amplitude error when there is no phase error and vice versa for MF and ZF precoders with 20, 100, and 500 antenna elements, respectively. The results show that the ZF precoder is more sensitive to errors than MF. It can also be seen that the impact of errors on the relative sum rate decreases as the number of antenna elements is increased.
\section{Conclusions}
In this paper, a performance analysis of single-cell, downlink massive MIMO has been presented. The results show that the 3GPP SCM and ITU UMa channel models give similar performance predictions as an IID Rayleigh channel if the path loss is the same for all UEs. Including path loss differences between UEs gives a significant performance reduction for the MF precoder, whereas the ZF precoder only gets a minor degradation. 

Furthermore,  the impact of phase and amplitude errors on performance was analyzed. Analytical analysis, based on approximations which are validated by simulations, show that these errors give an SINR loss that is independent of the number of antenna elements for the MF precoder in an IID Rayleigh channel. However, with the Tx power normalization used in this paper, the system will asymptotically be noise limited. Therefore, for large arrays, this loss can be compensated for by an increase in Tx power. It was shown by simulations that the ZF precoder is more sensitive to phase and amplitude errors than the MF precoder. Some simple analytical approximations for SINR and sum rate have been derived for the IID Rayleigh channel model. These approximations show how different parameters impact system performance.

%\bibliography{IEEEabrv,refs}

%\begin{thebibliography}{1}
%\bibitem{Young}
%M.~Young, \emph{The Technical Writer's Handbook}, Mill Valley, CA: University Science, 1989.
%
%\end{thebibliography}

\end{document}